\documentclass[12pt]{article}
\usepackage{graphicx}%
\usepackage{dcolumn}%
\usepackage{bm}%
\usepackage{amssymb,amsmath,latexsym}

\everymath{\displaystyle}

\begin{document}

\title{The BBGKY Hierarchy and Fokker-Planck Equation for Many-Body Dissipative Randomly Driven Systems}

\author{O.Yu. Sliusarenko\thanks{aslusarenko@kipt.kharkov.ua}\\
\small National Science Center ``Kharkiv Institute of Physics and Technology''\\[-0.8ex]
\small 1 Akademichna Str., 61108 Kharkiv, Ukraine\\
\and
A.V. Chechkin\\
\small National Science Center ``Kharkiv Institute of Physics and Technology''\\[-0.8ex]
\small 1 Akademichna Str., 61108 Kharkiv, Ukraine\\
\small Max-Planck Institute for Physics of Complex Systems\\[-0.8ex]
\small 38 Noethnitzer Strasse, 01187 Dresden, Germany\\
\and
Yu.V. Slyusarenko\\
\small National Science Center ``Kharkiv Institute of Physics and Technology''\\[-0.8ex]
\small 1 Akademichna Str., 61108 Kharkiv, Ukraine\\
\small Karazin National University, 4 Svobody Sq.\\[-0.8ex]
\small 61077 Kharkiv, Ukraine\\
}

\begin{abstract}
By generalizing Bogolyubov's reduced description method, we suggest a formalism to derive 
kinetic equations for many-body dissipative systems in external stochastic field. As a starting point, we use a  
stochastic Liouville equation obtained from Hamilton's equations taking dissipation and stochastic perturbations 
into account. The Liouville equation is then averaged over realizations of the stochastic field by an extension of the Furutsu-Novikov formula 
to the case of a non-Gaussian field. As the result, a generalization 
of the classical Bogolyubov-Born-Green-Kirkwood-Yvon (BBGKY) hierarchy
is derived. In order to get a kinetic equation 
for the one-particle distribution function, we use a regular cut off procedure of the BBGKY hierarchy by assuming weak interaction 
between the particles and weak intensity of the field. Within this approximation we get the corresponding Fokker-Planck equation 
for the system in a non-Gaussian stochastic field. Two particular cases by assuming
either Gaussian statistics of external perturbation or homogeneity of the
system are discussed.

\end{abstract}



\maketitle

\section{\label{sec.1}Introduction}

The research topic of the present paper has quite a long history, dating back to a rigorous justification 
of the Boltzmann kinetic equation from the microscopic point of view. A consistent dynamical foundation 
of statistical mechanics for classical (non-quantum) many-body systems has been given by Bogolyubov \cite{Bogolyubov}. 
His method of a reduced description allows to construct a systematic procedure for obtaining closed dissipative 
kinetic equations based on the Bogolyubov-Born-Green-Kirkwood-Yvon (BBGKY) chain of reversible equations 
for many-particle distribution functions. Bogolyubov's “functional hypothesis” reflects the idea 
of a relaxation time hierarchy. It allows one to truncate the BBGKY chain and get closed kinetic equations 
in two important cases, namely either low particle density, or weak particle interaction. In the former case 
we arrive at the Boltzmann equation, whereas the weak interaction
approximation leads to the Fokker-Planck equation \cite{AkhPel}. The BBGKY theory, together with
Choh-Uhlenbeck's generalization of the Boltzmann equation \cite{Uhlenbeck}, constitute a "cradle" of modern 
kinetic theory \cite{Ernst},
having a great impact on the further development of important concepts in the theory of condensed matter
and electromagnetic processes, see, e.g., \cite{Ernst, Cohen50, klim1}. For quantum systems a reduced 
description method was developed by Peletminskii and co-authors \cite{AkhPel}.

In the present paper we obtain a generalization 
of the classical BBGKY hierarchy by taking into account dissipative dynamics and non-Gaussian 
random driving. Apart from a fundamental scientific 
interest in developing and extending systematic methods of kinetic theory, 
our studies are motivated by the following physical reasons: 
Firstly, a wide variety of complex systems, 
such as colloidal suspensions, polymers, micelles etc., can be viewed as interacting Brownian particles that are 
in contact with a heat bath \cite{Russel, Edwards, Pokrovsky}. There are several 
approaches to the statistical description of Brownian dynamics where interaction between particles is taken into account, see, 
e.g., \cite{Rubi, Archer, Basu, HuiXia, Chavanis}, and references therein. 
In the case of a Gaussian external field, the system investigated 
in our paper can be considered as a 
``prototype'' of that of interacting Brownian particles. Secondly, there is a large and important class 
of many-body dissipative systems, such as granular media 
\cite{Brilliantov, Mehta} and their counterparts, traffic flows \cite{PRreview, RMPreview}. All these 
systems exhibit a reach phenomenology
and striking differences to the gases studied in the framework of the "standard" BBGKY theory \cite{Barrat}.
The kinetic approaches are under investigation, 
see \cite{Noije1998, Noije1999, Ernst2000, Noije2001, Pag, Maynar, Maynar_2, Khalil, Soria, Prados} 
for granular gas dynamics. We believe that alternative approaches 
and/or techniques are still
on the agenda because of the strong complexity of these systems. 
Thirdly, a large class of biological systems 
which are referred to as active matter is studied theoretically and experimentally with the use of methods 
of statistical physics \cite{Rama, Pavel, Marchetti}. In contrast to passive 
Brownian particles, 
active self-propelled objects have the ability to take up energy from the environment, to store it in an 
internal depot, and to convert internal energy into kinetic energy. 
The kinetic description of active Brownian dynamics requires taking into account both dissipative dynamics 
with active friction responsible for self-propulsion, and external active fluctuations that are correlated 
with the current direction of motion of an active particle \cite{Pavel}. The approach of the present paper 
may pave the way for constructing a consistent microscopic kinetic theory of active  gaseous-like systems.
We refer the reader to Refs.\cite{Pavel2, Ihle, Peshkov, Peshkov2, Ihle2} for recent 
developments and disputes in the field of the kinetic theory of active motion.

We would like to emphasize that the aim of the present paper is not to solve a particular problem 
of the systems mentioned above, but rather to outline a 
fundamental microscopic approach based on
the generalization of a well-known systematic method of classical kinetic theory.
However, in order to illustrate the simplest possible truncation of the BBGKY hierarchy, we consider 
a model of weakly interacting particles with the simplest 
form of dissipation. For such a model we obtain a Fokker-Planck equation and then 
present two particular cases by assuming either Gaussian statistics 
of external perturbations or homogeneity of the system under study. 

The paper is organized as follows. In Section \ref{sec.2}, starting 
from Hamilton's equations we introduce a stochastic Liouville equation 
for many-body dissipative systems in an external stochastic field. 
In Section \ref{sec.3} we describe an averaging procedure over realizations of the stochastic field. 
In Section \ref{sec.4} we get the analogue of the BBGKY chain. In Section \ref{sec.5} a closed Fokker-Planck
kinetic equation is obtained by assuming weak interaction between the particles. In Section \ref{sec.6} 
we deal with the simplest particular cases of the derived kinetic equation. 
Our results are summarized and discussed in Section \ref{sec.7}.

\section{\label{sec.2}Basic equations}

We consider a system consisting of $N$ identical particles of mass $m$, each of which is characterized 
by a spatial coordinate $\mathbf{x}_{\alpha } $, $1\le \alpha \le N$, measured from the center of mass, 
and momentum $\mathbf{p}_{\alpha } $, $1\le \alpha \le N$. We assume that the system is placed 
in an external stochastic field with potential $U^{\omega } \left(\mathbf{x},t\right)$ 
(here the index $\omega $ reflects the fact that $U^{\omega}\left(\mathbf{x},t\right)$ 
belongs to the set of random realizations of the external field). Following \cite{Archive}, 
we assume that the interaction between the particles consists of two parts, namely, ``reversible'' 
part described by the Hamiltonian $H$, and ``irreversible'' one, described by the dissipation function $R$. 
The Hamiltonian of the system reads:
\begin{equation}\label{eq__1_1_}
H=H_{0} +V=\sum _{1\le \alpha \le N}\left(\frac{\mathbf{p}_{\alpha }^{2} }{2m} +U^{\omega } \left(\mathbf{x}_{\alpha } ,t\right)\right) +\sum _{1\le \alpha <\beta \le N}V_{\alpha ,\beta },   
\end{equation}
where $V_{\alpha ,\beta } $ is the pair interaction potential,
\begin{equation}\label{eq__1_2_}
V_{\alpha ,\beta } \equiv V\left(\mathbf{x}_{\alpha \beta } \right), \mathbf{x}_{\alpha \beta } \equiv \mathbf{x}_{\alpha } -\mathbf{x}_{\beta } .
\end{equation}
In general case, we also assume that the dissipation function $R$ is determined only by the difference of coordinates $\mathbf{x}_{\alpha \beta } $ and momenta  $\mathbf{p}_{\alpha \beta } $ of the particles. In view of this, the system in the absence of external field exhibits the Galilean invariance. Thus, the dissipation function can be represented as follows:
\begin{equation}\label{eq__1_3_}
R=\sum _{1\le \alpha <\beta \le N}R_{\alpha ,\beta }, \qquad R_{\alpha ,\beta } \equiv R\left(\mathbf{x}_{\alpha \beta } ,\mathbf{p}_{\alpha \beta } \right),\qquad \mathbf{p}_{\alpha \beta } \equiv \mathbf{p}_{\alpha } -\mathbf{p}_{\beta } .
\end{equation}
Since the function $R$ is a scalar, it should depend only on $\mathbf{x}_{\alpha \beta }^{2} $, $\mathbf{p}_{\alpha \beta }^{2} $, $\mathbf{x}_{\alpha \beta } \mathbf{p}_{\alpha \beta }$
and their combinations.

In accordance with (\ref{eq__1_1_}) - (\ref{eq__1_3_}) the (generalized) Hamilton's equations can be written as:
\begin{equation}\label{eq__1_4_}
\dot{\mathbf{p}}_{\alpha } =-\frac{\partial H}{\partial \mathbf{x}_{\alpha } } -\frac{\partial R}{\partial \mathbf{p}_{\alpha } } , \qquad \dot{\mathbf{x}}_{\alpha } =\frac{\partial H}{\partial \mathbf{p}_{\alpha } } .
\end{equation}
Thus, the force $\mathbf{F}_{\alpha ,\beta } $, acting on the particle $\alpha $ from the particle $\beta $, is given by
\begin{equation} \label{eq__1_5_}
\mathbf{F}_{\alpha ,\beta } \equiv \mathbf{F}\left(\mathbf{x}_{\alpha \beta } ,\mathbf{p}_{\alpha \beta } \right)=-\frac{\partial V_{\alpha ,\beta } }{\partial \mathbf{x}_{\alpha } } -\frac{\partial R_{\alpha ,\beta } }{\partial \mathbf{p}_{\alpha } }  
\end{equation} 
Moreover, as it follows from (\ref{eq__1_1_}), the $\alpha $-th particle is under the influence of an external stochastic force $\mathbf{Y}_{\alpha }^{\omega } $,
\begin{equation}\label{eq__1_6_}
\mathbf{Y}_{\alpha }^{\omega } \equiv \mathbf{Y}^{\omega } \left(\mathbf{x}_{\alpha } ,t\right)=-\frac{\partial }{\partial \mathbf{x}_{\alpha } } U^{\omega } \left(\mathbf{x}_{\alpha } ,t\right).
\end{equation}

The time derivative of the total energy of the system in accordance with (\ref{eq__1_1_})
and (\ref{eq__1_4_}) is given by
\begin{equation}\label{eq__1_7_}
\frac{{\rm d} H}{{\rm d} t} =\frac{\partial '}{\partial t} \sum _{1\le \alpha \le N}U^{\omega } \left(\mathbf{x}_{\alpha } ,t\right) -\sum _{1\le \alpha \le N}\frac{\mathbf{p}_{\alpha } }{m}  \frac{\partial R}{\partial \mathbf{p}_{\alpha } } ,
\end{equation}
where the symbol ``prime'' in the partial time derivative means differentiation with respect to the explicit 
dependence of the potential $U^{\omega } \left(\mathbf{x}_{\alpha } ,t\right)$ on time. 

In this Section our aim is to obtain the Liouville equation. To this end, we first represent  (\ref{eq__1_4_}) in the equivalent form
\begin{equation}\label{eq__1_10_}
\dot{x}_{\alpha } \left(t\right)={h}_{\alpha }^{\omega } \left(x_{1} \left(t\right),...,x_{N} \left(t\right)\right), 1\le \alpha \le N,
\end{equation}
where we introduce the notation
\begin{equation}\label{eq__1_11_}
x_{a} \left(t\right)\equiv \left(\mathbf{x}_{a} \left(t\right), \mathbf{p}_{a} \left(t\right)\right).
\end{equation}
Alternatively, (\ref{eq__1_10_}) can be written as
\begin{eqnarray}\label{eq__1_12_}
\nonumber \dot{\mathbf{x}}_{\alpha } \left(t\right)=\mathbf{h}_{x\alpha }^{\omega } \left(x\left(t\right)\right),\\ 
\dot{\mathbf{p}}_{\alpha } \left(t\right)=\mathbf{h}_{p\alpha }^{\omega } \left(x\left(t\right)\right),
\end{eqnarray}
where
\begin{eqnarray}\label{eq__1_13_}
\nonumber \mathbf{h}_{x\alpha }^{\omega } \left(x\left(t\right)\right)=\frac{\partial H}{\partial \mathbf{p}_{\alpha } },\\
\mathbf{h}_{p\alpha }^{\omega } \left(x\left(t\right)\right)=-\frac{\partial H}{\partial \mathbf{x}_{\alpha } } -\frac{\partial R}{\partial \mathbf{p}_{\alpha } }.
\end{eqnarray}
The coordinate and momentum of the $\alpha $-th particle at time $t$ depend on coordinates and momenta $x_{0} \equiv \left(x_{1} \left(0\right),...,x_{N} \left(0\right)\right)$ of all the particles at initial time $t=0$,
\begin{equation}\label{eq__1_14_}
{x}_{\alpha }^{\omega } \left(t\right)={X}_{\alpha }^{\omega } \left(t,x_{0} \right)\equiv \left(\mathbf{X}_{\alpha }^{\omega } \left(t,x_{0} \right),\; \mathbf{P}_{\alpha }^{\omega } \left(t,x_{0} \right)\right),
\end{equation}
where the functions $\mathbf{X}_{\alpha }^{\omega } \left(t,x_{0} \right)$, $\; \mathbf{P}_{\alpha }^{\omega } \left(t,x_{0} \right)$ satisfy the Hamilton's equations (\ref{eq__1_4_}) 
(or (\ref{eq__1_10_})-(\ref{eq__1_13_})). Let us assume that at $t=0$ the initial conditions $x_{0} \equiv \left(x_{1} \left(0\right),...,x_{N} \left(0\right)\right)$ are distributed with the probability density $D\left(x_{1} \left(0\right),...,x_{N} \left(0\right);0\right)$.
Then
\begin{equation}\label{eq__1_15_}
\int  {\rm d} x_{1} \left(0\right)...\rm \mathbf{x}_{N} \left(0\right)D\left(x_{1} \left(0\right),...,x_{N} \left(0\right);0\right)\equiv \int  {\rm d} \mathbf{x}_{0} D\left(x_{0} ;0\right) =1.
\end{equation}
Then, at time $t$ the $N$-particle probability density $D^{\omega } \left(x_{1} ,...,x_{N} ;t\right)\equiv D^{\omega } \left(x;t\right)$, $x\equiv \left(x_{1} ,...,x_{N} \right)$, is determined by the expression
\begin{equation}\label{eq__1_16_}
D^{\omega } \left(x_{1} ,...,x_{N} ;t\right)=\int {\rm d} \mathbf{x}_{0} D\left(x_{0} ;0\right) \mathop{\prod }\limits_{1\le \alpha \le N} \delta \left({x}_{\alpha } -{X}_{\alpha }^{\omega } \left(t,x_{0} \right)\right).
\end{equation}

In \cite{Laskin} the stochastic Liouville equation is obtained for many-body system of non-interacting particles in external stochastic field. In \cite{Archive}, a similar procedure is used to obtain the Liouville equation for dissipative many-body system in the absence of a stochastic field. By directly combining these two procedures we arrive at the stochastic Liouville equation
for many-body dissipative system driven by external stochastic field,
\begin{equation}\label{eq__1_17_}
\frac{\partial D^{\omega } }{\partial t} +\sum _{1\le \alpha \le N}\frac{\partial }{\partial {x}_{\alpha } }  \left(D^{\omega } {h}_{\alpha }^{\omega } \right)=0,
\end{equation}
where the function ${h}_{\alpha }^{\omega } \left(x\left(t\right)\right)$ is given by the expressions (\ref{eq__1_12_}), (\ref{eq__1_13_}). The Liouville equation can be written in the
equivalent form as
\begin{equation}\label{eq__1_18_}
\frac{\partial D^{\omega } }{\partial t} -\left\{H,D^{\omega } \right\}=\sum _{1\le \alpha \le N}\frac{\partial }{\partial \mathbf{p}_{\alpha } }  \left(D^{\omega } \frac{\partial R}{\partial \mathbf{p}_{\alpha } } \right),
\end{equation}
where $\left\{A,B\right\}$ represent the $N$-particle Poisson brackets
\begin{equation}\label{eq__1_19_}
\left\{A,B\right\}\equiv \sum _{1\le \alpha \le N}\left(\frac{\partial A}{\partial \mathbf{x}_{\alpha } } \frac{\partial B}{\partial \mathbf{p}_{\alpha } } -\frac{\partial A}{\partial \mathbf{p}_{\alpha } } \frac{\partial B}{\partial \mathbf{x}_{\alpha } } \right) .
\end{equation}
For our purpose the following form of the Liouville equation appears to be more convenient:
\begin{eqnarray}\label{eq__1_20_}
 \frac{\partial D^{\omega } }{\partial t} +\sum _{1\le \alpha \le N}\frac{\mathbf{p}_{\alpha } }{m} \frac{\partial D^{\omega } }{\partial \mathbf{x}_{\alpha } }   +\sum _{1\le \alpha <\beta \le N}\frac{\partial }{\partial \mathbf{p}_{\alpha } } D^{\omega } \mathbf{F}_{\alpha ,\beta } 
 + \sum _{1\le \alpha \le N}\frac{\partial }{\partial \mathbf{p}_{\alpha } } D^{\omega } \mathbf{Y}_{\alpha }^{\omega }  =0,
\end{eqnarray}
where the quantities $\mathbf{F}_{\alpha ,\beta } $, $\mathbf{Y}_{\alpha }^{\omega } $ are defined by (\ref{eq__1_5_}), (\ref{eq__1_6_}). Equation (\ref{eq__1_20_}) is a typical example of the evolution equation with multiplicative noise. This circumstance poses the task of averaging over random realizations of the field $\mathbf{Y}_{\alpha }^{\omega } $.

\section{\label{sec.3}Averaging the stochastic Liouville equation}

We introduce the $N$-particle PDF $D\left(x_{1} ,...,x_{N} ;t\right)$ as a statistical average
of the $N$-particle PDF $D^{\omega } \left(x_{1} ,...,x_{N} ;t\right)$ (see (\ref{eq__1_16_})) over the stochastic external field $\mathbf{Y}^{\omega } \left(\mathbf{x},t\right)$ with the probability density $W[\mathbf{Y}^{\omega } ]$,
\begin{eqnarray} \label{eq__2_1_} 
\nonumber D\left(x_{1} ,...,x_{N} ;t\right)&&\equiv \left\langle D^{\omega } \left(x_{1} ,...,x_{N} ;t\right)\right\rangle _{\omega } ,  \\ 
\left\langle ...\right\rangle _{\omega }&&\equiv  \int D \mathbf{Y}^{\omega } \left(\mathbf{x},t\right)W[\mathbf{Y}^{\omega } ]...   
\end{eqnarray} 
After averaging of (\ref{eq__1_20_}) we get
\begin{eqnarray}\label{eq__2_2_}
 \frac{\partial D}{\partial t} +\sum _{1\le \alpha \le N}\frac{\mathbf{p}_{\alpha } }{m} \frac{\partial D}{\partial \mathbf{x}_{\alpha } }   +\sum _{1\le \alpha <\beta \le N}\frac{\partial }{\partial \mathbf{p}_{\alpha } } D \mathbf{F}_{\alpha ,\beta } 
 +\sum _{1\le \alpha \le N}\frac{\partial }{\partial \mathbf{p}_{\alpha } } \left\langle D^{\omega } \mathbf{Y}_{\alpha }^{\omega } \right\rangle _{\omega }  =0.
\end{eqnarray}
To get a closed equation for the $N$-particle PDF we need to express the quantity
$\left\langle D^{\omega } \mathbf{Y}_{\alpha }^{\omega } \right\rangle _{\omega } $ through $D\left(x_{1} ,...,x_{N} ;t\right)$. To this end, we use the Furutsu-Novikov formula \cite{Furutsu, Novikov}, which was generalized in \cite{nonGaussian} to the case of a non-Gaussian stochastic field.  Below we follow the procedure of that paper and present some details of calculations for the
readers' convenience. At first we introduce the moments\textbf{ ${Y}_{i_{1} ...i_{n} } \left(\mathbf{x}_{1} ,...,\mathbf{x}_{n} ;t_{1} ,...,t_{n} \right)$ } of the stochastic field $\mathbf{Y}^{\omega } \left(\mathbf{x},t\right)$, 
\begin{eqnarray}\label{eq__2_3_}
\nonumber {Y}_{i_{1} ...i_{n} } \left(\mathbf{x}_{1} ,...,\mathbf{x}_{n} ;t_{1} ,...,t_{n} \right)&&\equiv \left\langle {Y}_{i_{1} }^{\omega } \left(\mathbf{x}_{1} ,t_{1} \right)...{Y}_{i_{n} }^{\omega } \left(\mathbf{x}_{n} ,t_{n} \right)\right\rangle _{\omega } \\
{Y}_{i}^{\omega } \left(\mathbf{x},t\right)&&\equiv \left(\mathbf{Y}^{\omega } \left(\mathbf{x},t\right)\right)_{i} , \qquad i=\left(1,2,3\right).
\end{eqnarray}
The generating functional $P\left(v;{Y}_{a} \right)$ of these moments is determined by
\begin{equation}\label{eq__2_4_}
P\left(v;{Y}_{a} \right)\equiv \left\langle \exp \left(\int {\rm d}\mathbf{x} \int _{-\infty }^{\infty } {\rm d} tv_{i} \left(\mathbf{x},t\right) {Y}_{i}^{\omega } \left(\mathbf{x},t\right)\right)\right\rangle _{\omega } ,
\end{equation}
where $v_{i} \left(\mathbf{x},t\right)$ is a functional argument, and summation is implied over the repeated indices. By ${Y}_{a} $ in the left-hand side of (\ref{eq__2_4_}) we denote the entire set of the moments ${Y}_{i_{1} ...i_{n} } \left(\mathbf{x}_{1} ,...,\mathbf{x}_{n} ;t_{1} ,...,t_{n} \right)$. Along with the moments (\ref{eq__2_3_}) we introduce the correlation functions $y_{i_1...i_s}\left({\bf x}_1,...,{\bf x}_s; t_1,...,t_s\right)$,
\begin{eqnarray}\label{CorFunc}
\nonumber {Y_i}\left( {{\bf{x}};t} \right) &&\equiv {y_i}\left( {{\bf{x}};t} \right), \\
\nonumber {Y_{{i_1}{i_2}}}\left( {{{\bf{x}}_1},{{\bf{x}}_2};{t_1},{t_2}} \right) &&= {Y_{{i_1}}}\left( {{{\bf{x}}_1};{t_1}} \right){Y_{{i_2}}}\left( {{{\bf{x}}_2};{t_2}} \right) + {y_{{i_1}{i_2}}}\left( {{{\bf{x}}_1},{{\bf{x}}_2};{t_1},{t_2}} \right), \\
\nonumber {Y_{{i_1}{i_2}{i_3}}}\left( {{{\bf{x}}_1},{{\bf{x}}_2},{{\bf{x}}_3};{t_1},{t_2},{t_3}} \right) &&= {Y_{{i_1}}}\left( {{{\bf{x}}_1};{t_1}} \right){Y_{{i_2}}}\left( {{{\bf{x}}_2};{t_2}} \right){Y_{{i_3}}}\left( {{{\bf{x}}_3};{t_3}} \right) \\ 
\nonumber && + {Y_{{i_1}}}\left( {{{\bf{x}}_1};{t_1}} \right){y_{{i_2}{i_3}}}\left( {{{\bf{x}}_2},{{\bf{x}}_3};{t_2},{t_3}} \right) + {Y_{{i_2}}}\left( {{{\bf{x}}_2};{t_2}} \right){y_{{i_1}{i_3}}}\left( {{{\bf{x}}_1},{{\bf{x}}_3};{t_1},{t_3}} \right) \\
&& + {Y_{{i_3}}}\left( {{{\bf{x}}_3};{t_3}} \right){y_{{i_1}{i_2}}}\left( {{{\bf{x}}_1},{{\bf{x}}_2};{t_1},{t_2}} \right) + {y_{{i_1}{i_2}{i_3}}}\left( {{{\bf{x}}_1},{{\bf{x}}_2},{{\bf{x}}_3};{t_1},{t_2},{t_3}} \right).
\end{eqnarray}

The generating functional of the correlation functions reads
\begin{eqnarray} \label{eq__2_5_}
\nonumber {\cal{P}}\left(v;y_{a} \right)  \equiv  \sum _{n=2}^{\infty }\frac{1}{n!} \int {\rm d}\mathbf{x}_{1}  \int _{-\infty }^{\infty } {\rm d} t_{1} ...\int {\rm d}\mathbf{x}_{n}  \int _{-\infty }^{\infty } {\rm d} t_{n} \, v_{i_{1} } \left(\mathbf{x}_{1} ,t_{S} \right) ...v_{i_{n} } \left(\mathbf{x}_{n} ,t_{n} \right) \\
\times  y_{i_{1} ...i_{n} } \left(\mathbf{x}_{1} ,...,\mathbf{x}_{n} ;t_{1} ,...,t_{n} \right) ,
\end{eqnarray}
and it is related to the generating functional $P\left(v;\mathbf{Y}_{a} \right)$ by the following
formula:
\begin{equation}\label{eq__2_6_}
P\left(v;{Y}_{a} \right)=\exp \left(\int {\rm d}\mathbf{x} \int _{-\infty }^{\infty } {\rm d} t \,v_{i} \left(\mathbf{x},t\right) {Y}_{i} \left(\mathbf{x},t\right)\right)\exp \left\{{\cal{P}}\left(v;y_{a} \right)\right\},
\end{equation}
where 
\begin{equation} \label{eq__2_7_}
\mathbf{Y}\left(\mathbf{x},t\right)\equiv \left\langle \mathbf{Y}^{\omega } \left(\mathbf{x},t\right)\right\rangle _{\omega } , 
\end{equation} 
and by $y_{a} $ we denote the entire set of the correlation functions $y_{i_{1} ...i_{n} } \left(\mathbf{x}_{1} ,...,\mathbf{x}_{n} ;t_{1} ,...,t_{n} \right)$. It was shown in \cite{nonGaussian} that the result of the averaging of an arbitrary functional $A[\mathbf{Y}^{\omega } \left(\mathbf{x},t\right)]$ over an external stochastic field with distribution $W[\mathbf{Y}^{\omega } \left(\mathbf{x},t\right)]$ can be presented as
\begin{equation}\label{eq__2_8_}
\left\langle A[\mathbf{Y}^{\omega } \left(\mathbf{x},t\right)]\right\rangle _{\omega } =\exp \left\{{\cal{P}}\left(\frac{\delta }{\delta Y} ;y_{a} \right)\right\}A[Y],
\end{equation}
where ${\cal{P}}\left(\frac{\delta }{\delta Y} ;y_{a} \right)$ is the generating functional (\ref{eq__2_5_}), in which the functional argument $v_{i} \left(\mathbf{x},t\right)$ is replaced by the operation of the functional differentiation over ${Y}_{i} \left(\mathbf{x},t\right)$ (see (\ref{eq__2_3_}), (\ref{eq__2_7_})),
\begin{equation}\label{eq__2_9_}
v_{i} \left(\mathbf{x},t\right)\to \delta /\delta {Y}_{i} \left(\mathbf{x},t\right).
\end{equation}
As the consequence of (\ref{eq__2_8_}),
\begin{eqnarray*} 
\nonumber \left\langle {Y}_{i}^{\omega } \left(\mathbf{x},t\right)A[\mathbf{Y}^{\omega } \left(\mathbf{x},t\right)]\right\rangle _{\omega } && =\left({Y}_{i} \left(\mathbf{x},t\right)+\frac{\delta {\cal{P}}\left(v;y_{a} \right)}{\delta v_{i} \left(\mathbf{x},t\right)} \right)_{v_{i} \to \delta /\delta {Y}_{i} }\\
&& \times  \exp \left\{{\cal{P}}\left(\frac{\delta }{\delta Y} ;y_{a} \right)\right\}A[Y].
\end{eqnarray*}
The same formula can be rewritten as
\begin{eqnarray} \label{eq__2_10_}
\nonumber \left\langle {Y}_{i}^{\omega } \left(\mathbf{x},t\right)A[\mathbf{Y}^{\omega } \left(x,t\right)]\right\rangle _{\omega } && ={Y}_{i} \left(\mathbf{x},t\right)\left\langle A[\mathbf{Y}^{\omega } \left(\mathbf{x},t\right)]\right\rangle _{\omega }\\
&& + \left\langle \left. \frac{\delta {\cal{P}}\left(v;y_{a} \right)}{\delta v_{i} \left(\mathbf{x},t\right)} \right|_{v_{i} \to \delta /\delta Y_{i} } A[\mathbf{Y}^{\omega } \left(\mathbf{x},t\right)]\right\rangle _{\omega } .
\end{eqnarray}
Equation (\ref{eq__2_10_}) is, indeed, the generalization of the Furutsu-Novikov formula to an arbitrary distribution of the external stochastic field (of course, it is assumed that this distribution has moments of any order). Note, that for the Gaussian distribution of the external stochastic field (\ref{eq__2_10_}) takes the form
\begin{eqnarray} \label{eq__2_11_}
\nonumber \left\langle Y_{i}^{\omega } \left(\mathbf{x},t\right)A[\mathbf{Y}^{\omega } ]\right\rangle _{\omega }  =Y_{i} \left(\mathbf{x},t\right)\left\langle A[\mathbf{Y}^{\omega } ]\right\rangle _{\omega } +\int {\rm d}\mathbf{x}' \int _{-\infty }^{\infty } {\rm d} t' \, y_{ij} \left(\mathbf{x},\mathbf{x}',t-t'\right)\\
  \times  \left\langle \frac{\delta A[\mathbf{Y}^{\omega } ]}{\delta {Y}_{j} \left(\mathbf{x}',t'\right)} \right\rangle _{\omega } ,
\end{eqnarray}
where $y_{ij} \left(\mathbf{x},\mathbf{x}',t-t'\right)$ is the pair correlation function of the external Gaussian noise,
\begin{equation*}
y_{ij} \left(\mathbf{x},\mathbf{x}',t-t'\right)=\left\langle Y_{i}^{\omega } \left(\mathbf{x},t\right)Y_{j}^{\omega } \left(\mathbf{x}',t'\right)\right\rangle _{\omega } -\left\langle Y_{i}^{\omega } \left(\mathbf{x},t\right)\right\rangle _{\omega } \left\langle Y_{j}^{\omega } \left(\mathbf{x}',t'\right)\right\rangle _{\omega } .   
\end{equation*}
When $Y_{i} \left(\mathbf{x},t\right)\equiv 0$ expression (\ref{eq__2_11_}) coincides with that obtained in \cite{Furutsu, Novikov}.

We now use the generalized Furutsu-Novikov formula to calculate the last term on the left-hand side of equation (\ref{eq__2_2_}). Indeed, there is a functional dependence of the distribution function $D^{\omega } \left(x_{1} ,...,x_{N} ;t\right)$ on the external stochastic field $Y_{i}^{\omega } \left(\mathbf{x},t\right)$,
\begin{equation*}
D^{\omega } \equiv D^{\omega } [\mathbf{Y}^{\omega } ],
\end{equation*}
as is evident from (\ref{eq__1_16_}) and (\ref{eq__1_20_}). According to  (\ref{eq__2_10_}), the mean value $\left\langle D^{\omega } \mathbf{Y}^{\omega } \right\rangle _{\omega } $ can be represented as
\begin{equation}\label{eq__2_12_}
\left\langle D^{\omega } [\mathbf{Y}^{\omega } ]Y_{i}^{\omega } \left(\mathbf{x},t\right)\right\rangle _{\omega } =Y_{i} \left(\mathbf{x},t\right)D+\left\langle \left. \frac{\delta {\cal{P}}\left(v;y_{a} \right)}{\delta v_{i} \left(\mathbf{x},t\right)} \right|_{v_{i} \to \delta /\delta Y_{i} } D^{\omega } [\mathbf{Y}^{\omega } ]\right\rangle _{\omega } .
\end{equation}
Note that, in accordance with (\ref{eq__2_5_})
\begin{eqnarray} \label{eq__2_13_}
\nonumber \frac{\delta {\cal{P}}\left(v;y_{a} \right)}{\delta v_{i} \left(\mathbf{x},t\right)}  =\sum _{n=1}^{\infty }\frac{1}{n!} \int  {\rm d} \mathbf{x}_{1}  \int _{-\infty }^{\infty } {\rm d} t_{1} ...\int  {\rm d} \mathbf{x}_{n}  \int _{-\infty }^{\infty } {\rm d} t_{n} \, v_{i_{1} } \left(\mathbf{x}_{1} ,t_{S} \right) ...v_{i_{n} } \left(\mathbf{x}_{n} ,t_{n} \right) \\
  \times  y_{ii_{1} ...i_{n} } \left(\mathbf{x},\mathbf{x}_{1} ,...,\mathbf{x}_{n} ;t,t_{1} ,...,t_{n} \right),
\end{eqnarray}
therefore, the correlator $\left\langle \left. \frac{\delta {\cal{P}}\left(v;y_{a} \right)}{\delta v_{i} \left(\mathbf{x},t\right)} \right|_{v_{i} \to \delta /\delta Y_{i} } D^{\omega } [\mathbf{Y}^{\omega } ]\right\rangle _{\omega } $ in (\ref{eq__2_12_}) reads as
\begin{eqnarray}\label{eq__2_14_}
\nonumber \left\langle \left. \frac{\delta {\cal{P}}\left(v;y_{a} \right)}{\delta v_{i} \left(\mathbf{x},t\right)} \right|_{v_{i} \to \delta /\delta Y_{i} } D^{\omega } [\mathbf{Y}^{\omega } ]\right\rangle _{\omega } &&= \\
\nonumber \sum _{n=1}^{\infty }\frac{1}{n!} \int  {\rm d} \mathbf{x}_{1}  \int _{-\infty }^{\infty } &&{\rm d} t_{1} ...\int  {\rm d} \mathbf{x}_{n}  \int _{-\infty }^{\infty } {\rm d} t_{n} \,    y_{ii_{1} ...i_{n} } \left(\mathbf{x},\mathbf{x}_{1} ,...,\mathbf{x}_{n} ;t,t_{1} ,...,t_{n} \right)\\
&&  \times \left\langle \frac{\delta ^{n} D^{\omega } [\mathbf{Y}^{\omega } ]}{\delta {Y}_{i_{1} }^{\omega } \left(\mathbf{x}_{1} ,t_{1} \right)...\delta {Y}_{i_{n} }^{\omega } \left(\mathbf{x}_{n} ,t_{n} \right)} \right\rangle _{\omega } .
\end{eqnarray}

To go further, we consider the first term of the sum in the right-hand side of (\ref{eq__2_14_}),
\begin{equation}\label{eq__2_15_}
I_{i} \equiv \int  {\rm d} \mathbf{x}' \int _{-\infty }^{\infty } {\rm d} t' \, y_{ij} \left(\mathbf{x},\mathbf{x}';t,t'\right)\left\langle \frac{\delta D^{\omega } [\mathbf{Y}^{\omega } ]}{\delta Y_{j}^{\omega } \left(\mathbf{x}',t'\right)} \right\rangle _{\omega }.
\end{equation}

We assume that the pair correlation function $y_{ij} \left(\mathbf{x},\mathbf{x}';t,t'\right)$ has a sharp peak at $t\approx t'$ and decays rapidly within the interval $\left|t-t'\right|\le \tau _{0} $, where $\tau _{0} $ is a typical correlation time of the stochastic field. Then it is enough to calculate the functional derivative $\frac{\delta D^{\omega } [\mathbf{Y}^{\omega } ]}{\delta Y_{j}^{\omega } \left(\mathbf{x}',t'\right)} $ only at $t\approx t'$.
The variational derivative $\frac{\delta D^{\omega } [\mathbf{Y}^{\omega } ]}{\delta Y_{j}^{\omega } \left(\mathbf{x}',t'\right)} $ at $t\approx t'$ undergoes a jump,
\begin{equation}\label{eq__2_16_}
\frac{\delta D^{\omega } [\mathbf{Y}^{\omega } ]}{\delta Y_{j}^{\omega } \left(\mathbf{x}',t'\right)} \ne 0,\quad t'\le t, \quad \frac{\delta D^{\omega } [\mathbf{Y}^{\omega } ]}{\delta Y_{j}^{\omega } \left(\mathbf{x}',t'\right)} =0,\quad t'>t.
\end{equation}
Indeed, according to (\ref{eq__1_20_}), the function $D^{\omega } \left(t\right)$ cannot depend on the field $Y_{j}^{\omega } \left(\mathbf{x}',t'\right)$ taken at later time instant $t'$.  Therefore, the integration over $t'$ in (\ref{eq__2_15_}) is performed in the range from $-\infty $ to $t$, not from  $-\infty $ to $+\infty $.

Differentiating (\ref{eq__1_20_}) with respect to $Y_{j}^{\omega } \left(\mathbf{x}',t'\right)$ and noting that, according to (\ref{eq__2_16_}), the derivative $\frac{\partial }{\partial t} \frac{\delta D^{\omega } [\mathbf{Y}^{\omega } ]}{\delta Y_{j}^{\omega } \left(\mathbf{x}',t'\right)} $ must contain a $\delta $- function singularity in time (while the functional derivative $\frac{\delta D^{\omega } [\mathbf{Y}^{\omega } ]}{\delta Y_{j}^{\omega } \left(\mathbf{x}',t'\right)} $ does not), it is easy to obtain the following expression (see also  \cite{nonGaussian}):
\begin{equation}\label{eq__2_17_}
\frac{\delta D^{\omega } [\mathbf{Y}^{\omega } ]}{\delta Y_{j}^{\omega } \left(\mathbf{x}',t'\right)} =-\vartheta \left(t-t'\right)\sum _{1\le \alpha \le N}\delta \left(\mathbf{x}'-\mathbf{x}_{\alpha } \right)\frac{\partial D^{\omega } [\mathbf{Y}^{\omega } ]}{\partial p_{\alpha j} }.
\end{equation}
This formula allows us to present the quantity $I_{i} $, (\ref{eq__2_15_}), as follows:
\begin{equation}\label{eq__2_18_}
I_{i} =\int _{-\infty }^{t} {\rm d} t' \sum _{1\le \alpha \le N}y_{ij} \left(\mathbf{x},\mathbf{x}_{\alpha } ;t,t'\right)\frac{\partial D}{\partial p_{\alpha j} }  .
\end{equation}
For the Gaussian processes, (\ref{eq__2_18_}) is sufficient to evaluate expression (\ref{eq__2_12_}), and hence to derive in a final form the averaged Liouville equation for dissipative system of many particles in an external stochastic field,
\begin{eqnarray} \label{eq__2_19_}
\nonumber  \frac{\partial D}{\partial t}  +\sum _{1\le \alpha \le N}\frac{\mathbf{p}_{\alpha } }{m} \frac{\partial D}{\partial \mathbf{x}_{\alpha } } && +\sum _{1\le \alpha <\beta \le N}\frac{\partial D\mathbf{F}_{\alpha ,\beta } }{\partial \mathbf{p}_{\alpha } }  +\sum _{1\le \alpha \le N}\mathbf{Y}\left(\mathbf{x}_{\alpha } ,t\right)\frac{\partial D}{\partial \mathbf{p}_{\alpha } }  \\
&& -\int _{-\infty }^{t} {\rm d} t' \sum _{1\le \alpha \le N}y_{ij} \left(\mathbf{x}_{\alpha } ,\mathbf{x}_{\beta } ;t,t'\right)\frac{\partial ^{2} D}{\partial p_{\alpha i} \partial p_{\beta j} }  =0.
\end{eqnarray}
In fact, the method used here also allows for obtaining the averaged Liouville equation in case of non-Gaussian distribution of the external stochastic field (remind, that we assume existence
of the moments of all orders). To do this, according to (\ref{eq__2_14_}), one should calculate the functional derivative of $n$-th order. This quantity can be obtained with the help of  (\ref{eq__2_17_}). Indeed, by differentiating (\ref{eq__2_17_}) over ${Y}_{l}^{\omega } \left(\mathbf{x}'',t''\right)$ we get
\begin{equation*}
\frac{\delta ^{2} D^{\omega } [\mathbf{Y}^{\omega } ]}{\delta Y_{j}^{\omega } \left(\mathbf{x}',t'\right)\delta {Y}_{l}^{\omega } \left(\mathbf{x}'',t''\right)} =-\vartheta \left(t-t'\right)\sum _{1\le \alpha \le N}\delta \left(\mathbf{x}'-\mathbf{x}_{\alpha } \right)\frac{\partial }{\partial p_{\alpha j} }  \frac{\delta D^{\omega } [\mathbf{Y}^{\omega } ]}{\delta Y_{l}^{\omega } \left(\mathbf{x}'',t''\right)} .
\end{equation*}
Next, we use (\ref{eq__2_17_}) again,
\begin{eqnarray*} 
\nonumber \frac{\delta ^{2} D^{\omega } [\mathbf{Y}^{\omega } ]}{\delta Y_{j}^{\omega } \left(\mathbf{x}',t'\right)\delta Y_{l}^{\omega } \left(\mathbf{x}'',t''\right)} && =  \vartheta \left(t-t'\right)\vartheta \left(t-t''\right)\\
 && \times  \sum _{1\le \alpha \le N}\sum _{1\le \beta \le N}\delta \left(\mathbf{x}'-\mathbf{x}_{\alpha } \right)\delta \left(\mathbf{x}''-\mathbf{x}_{\beta } \right)\frac{\partial ^{2} D^{\omega } [\mathbf{Y}^{\omega } ]}{\partial p_{\alpha j} \partial p_{\beta l} }   .
\end{eqnarray*}
Repeating this procedure, it is easy to come to the following expression:
\begin{eqnarray}\label{eq__2_20_}
\nonumber &\frac{\delta ^{n} D^{\omega } [\mathbf{Y}^{\omega } ]}{\delta {Y}_{i_{1} }^{\omega } \left(\mathbf{x}_{1} ,t_{1} \right)...\delta {Y}_{i_{n} }^{\omega } \left(\mathbf{x}_{n} ,t_{n} \right)}& = \\
\nonumber &&\left(-1\right)^{n} \vartheta \left(t-t_{1} \right)...\vartheta \left(t-t_{n} \right)\left(\sum _{1\le \alpha _{1} \le N}\delta \left(\mathbf{x}_{1} -\mathbf{x}_{\alpha _{1} } \right)\frac{\partial }{\partial {p}_{\alpha _{1} i_{1} } }  \right)...\\
 &&\times  \left(\sum _{1\le \alpha _{n} \le N}\delta \left(\mathbf{x}_{n} -\mathbf{x}_{\alpha _{n} } \right)\frac{\partial }{\partial {p}_{\alpha _{n} i_{n} } }  \right)D^{\omega } [\mathbf{Y}^{\omega } ].
\end{eqnarray}
After substituting (\ref{eq__2_20_}) into (\ref{eq__2_14_}) and using  (\ref{eq__2_12_}), the evolution equation (\ref{eq__2_2_}) can be written as:
\begin{eqnarray}\label{eq__2_21_}
\nonumber \frac{\partial D}{\partial t} && +\sum _{1\le \alpha \le N}\frac{\mathbf{p}_{\alpha } }{m} \frac{\partial D}{\partial \mathbf{x}_{\alpha } }  +\sum _{1\le \alpha <\beta \le N}\frac{\partial D\mathbf{F}_{\alpha ,\beta } }{\partial \mathbf{p}_{\alpha } }  +\sum _{1\le \alpha \le N}\mathbf{Y}\left(\mathbf{x}_{\alpha } ,t\right)\frac{\partial D}{\partial \mathbf{p}_{\alpha } } \\
\nonumber && + \sum _{n=2}^{\infty }\left \{\frac{\left((-1\right)^{n-1} }{\left(n-1\right)!} \int _{-\infty }^{t} {\rm d} t_{2}  ...\int _{-\infty }^{t} {\rm d} t_{n}  \sum _{1\le \alpha _{1} \le N}...\right.\\
 && \times  \left. \sum _{1\le \alpha _{n} \le N}y_{i_{1} ...i_{n} } \left(\mathbf{x}_{\alpha _{1} } ,...,\mathbf{x}_{\alpha _{n} } ;t,t_{2} ,...,t_{n} \right) \frac{\partial ^{n} D}{\partial {p}_{\alpha _{1} i_{1} } ...\partial {p}_{\alpha _{n} i_{n} } }\right\} =0.
\end{eqnarray}
Equation (\ref{eq__2_21_}) is the generalized Liouville equation for dissipative systems of many particles, averaged over the external non-Gaussian stochastic field. 

In case of stationary stochastic field the Liouville equation (\ref{eq__2_21_}) can be further simplified. Indeed, the correlation functions $y_{i_{1} ...i_{n} } \left(\mathbf{x}_{1} ,...,\mathbf{x}_{n} ;t,t_{2} ,...,t_{n} \right)$ can be represented as
\begin{eqnarray} \label{}
 y_{i_{1} ...i_{n} } \left(\mathbf{x}_{\alpha _{1} } ,...,\mathbf{x}_{\alpha _{n} } ;t,t_{2} ,...,t_{n} \right)
 \equiv y_{i_{1} ...i_{n} } \left(\mathbf{x}_{\alpha _{1} } ,...,\mathbf{x}_{\alpha _{n} } ;\left|t_{2} -t\right|,...,\left|t_{n} -t\right|\right).
\end{eqnarray}
To simplify further calculations we also assume that the average value of the external stochastic field is zero,  
\begin{equation}\label{eq__2_23_}
\mathbf{Y}\left(\mathbf{x},t\right)\equiv \left\langle \mathbf{Y}^{\omega } \left(\mathbf{x},t\right)\right\rangle _{\omega } \equiv 0.
\end{equation}
Then the Lioville equation (\ref{eq__2_21_}) reads
\begin{eqnarray}\label{eq__2_24_}
\nonumber \frac{\partial D}{\partial t}  +\sum _{1\le \alpha \le N}\frac{\mathbf{p}_{\alpha } }{m} \frac{\partial D}{\partial \mathbf{x}_{\alpha } } && +\sum _{1\le \alpha <\beta \le N}\frac{\partial D\mathbf{F}_{\alpha ,\beta } }{\partial \mathbf{p}_{\alpha } }
 +\sum _{n=2}^{\infty }\frac{\left(-1\right)^{n-1} }{2^{n-1} \left(n-1\right)!}  \sum _{1\le \alpha _{1} \le N}... \\
&& \times \sum _{1\le \alpha _{n} \le N}\bar{y}_{i_{1} ...i_{n} } \left(\mathbf{x}_{\alpha _{1} } ,...,\mathbf{x}_{\alpha _{n} } \right) \frac{\partial ^{n} D}{\partial {p}_{\alpha _{1} i_{1} } ...\partial {p}_{\alpha _{n} i_{n} } } =0,
\end{eqnarray}
where the correlation functions 
\begin{equation} \label{correlation}
\bar{y}_{i_{1} ...i_{n} } \left(\mathbf{x}_{\alpha _{1} } ,...,\mathbf{x}_{\alpha _{n} } \right) \equiv \int _{-\infty }^{\infty } {\rm d} t_{2}  ...\int _{-\infty }^{\infty } {\rm d} t_{n} \,  y_{i_{1} ...i_{n} } \left(\mathbf{x}_{\alpha _{1} } ,...,\mathbf{x}_{\alpha _{n} } ;\left|t_{2} -t\right|,...,\left|t_{n} -t\right|\right).
\end{equation}
are introduced.

\section{\label{sec.4}Analogue of the BBGKY hierarchy for dissipative system in external stochastic field}

Along with the probability density $D$ one may introduce the probability density of finding one or more particles in given elements of the phase space, regardless of where the other particles are. These probabilities can be obtained by integrating the function $D$ over  all variables except those related to the considered particles,
\begin{eqnarray}\label{eq__3_1_}
\nonumber f_{S} \left({x}_{1} ,...,{x}_{S} ;t\right)=V^{S} \int  {\rm d} {x}_{S+1}  ...\int  {\rm d} {x}_{N} \, D\left({x}_{1} ,...,{x}_{N} ;t\right),\\
{x}_{\alpha } \equiv \left(\mathbf{x}_{\alpha } ,\mathbf{p}_{\alpha } \right),
\end{eqnarray}
where $D\left({x}_{1} ,...,{x}_{N} ;t\right)$ satisfies (\ref{eq__2_21_}) (or (\ref{eq__2_24_})) and $V$ is the system's volume. Following the procedure of \cite{Bogolyubov,AkhPel}, one gets the following equation for $S$- particle distribution function $f_{S} \left({x}_{1} ,...,{x}_{S} ;t\right)$:
\begin{eqnarray}\label{eq__3_2_}
\nonumber \frac{\partial f_{S} }{\partial t}  =&&-\sum _{1\le \alpha \le S}\frac{\mathbf{p}_{\alpha } }{m} \frac{\partial f_{S} }{\partial \mathbf{x}_{\alpha } }  -\sum _{1\le \alpha <\beta \le S}\frac{\partial f_{S} \mathbf{F}_{\alpha ,\beta } }{\partial \mathbf{p}_{\alpha } }  -\sum _{1\le \alpha \le N}\mathbf{Y}\left(\mathbf{x}_{\alpha } ,t\right)\frac{\partial f_{S} }{\partial \mathbf{p}_{\alpha } } \\
\nonumber && -\sum _{n=2}^{\infty }\frac{\left(-1\right)^{n-1} }{\left(n-1\right)!}  \int _{-\infty }^{t} {\rm d} t_{2}  ...\int _{-\infty }^{t} {\rm d} t_{n}  \sum _{1\le \alpha _{1} \le S}...\\
\nonumber && \times \sum _{1\le \alpha _{n} \le S}y_{i_{1} ...i_{n} } \left(\mathbf{x}_{\alpha _{1} } ,...,\mathbf{x}_{\alpha _{n} } ;t,t_{2} ,...,t_{n} \right) \frac{\partial ^{n} f_{S} }{\partial {p}_{\alpha _{1} i_{1} } ...\partial {p}_{\alpha _{n} i_{n} } } \\
&& -\frac{1}{v} \sum _{1\le \alpha \le S}\frac{\partial }{\partial \mathbf{p}_{\alpha } }  \int  {\rm d} {x}_{S+1} \, f_{S+1} \mathbf{F}_{\alpha ,S+1} , \qquad v\equiv \frac{V}{N} ,
\end{eqnarray}
if the function  $D\left({x}_{1} ,...,{x}_{N} ;t\right)$ satisfies (\ref{eq__2_21_}), and
\begin{eqnarray}\label{eq__3_3_}
\nonumber \frac{\partial f_{S} }{\partial t} = &&-\sum _{1\le \alpha \le S}\frac{\mathbf{p}_{\alpha } }{m} \frac{\partial f_{S} }{\partial \mathbf{x}_{\alpha } }  -\sum _{1\le \alpha <\beta \le S}\frac{\partial f_{S} \mathbf{F}_{\alpha ,\beta } }{\partial \mathbf{p}_{\alpha } }  \\
\nonumber && -\sum _{n=2}^{\infty }\frac{\left(-1\right)^{n-1} }{2^{n-1} \left(n-1\right)!}  \sum _{1\le \alpha _{1} \le S}... \sum _{1\le \alpha _{n} \le S}\bar{y}_{i_{1} ...i_{n} } \left(\mathbf{x}_{\alpha _{1} } ,...,\mathbf{x}_{\alpha _{n} } \right) \frac{\partial ^{n} f_{S} }{\partial {p}_{\alpha _{1} i_{1} } ...\partial {p}_{\alpha _{n} i_{n} } } \\
&& -\frac{1}{v} \sum _{1\le \alpha \le S}\frac{\partial }{\partial \mathbf{p}_{\alpha } }  \int  {\rm d} \mathbf{x}_{S+1} \, f_{S+1} \mathbf{F}_{\alpha ,S+1} , \qquad v\equiv \frac{V}{N} ,
\end{eqnarray}
if the function $D\left({x}_{1} ,...,{x}_{N} ;t\right)$ satisfies (\ref{eq__2_24_}). The force $\mathbf{F}_{\alpha ,\beta } $ in (\ref{eq__3_2_}) and (\ref{eq__3_3_}) is still defined by (\ref{eq__1_5_}). As it is seen, the equation for the $S$- particle distribution function contains an $S+1$- particle distribution function, that is, in fact, we have obtained endless chain of equations (\ref{eq__3_2_}) and (\ref{eq__3_3_}), which are generalizations of the famous BBGKY hierarchy for the dissipative systems of many particles under the influence of an external stochastic field. The chains (\ref{eq__3_2_}) and (\ref{eq__3_3_}) are equivalent to the Liouville equations (\ref{eq__2_21_}) or (\ref{eq__2_24_}), respectively, and thus provide the same level of complexity in the system's description.

In what follows we restrict ourselves to the case of a stationary stochastic field that is, we consider the chain (\ref{eq__2_24_}) and omit the bar in the notation of the correlation functions defined by (\ref{correlation}). The generalization to the non-stationary case given by (\ref{eq__3_2_}) is straightforward.

A significant simplification in the system's description occurs in two cases: when interaction 
between the particles is weak, or the particle density is low, and the interaction is arbitrary, 
but such that it does not lead to the formation of bound states \cite{AkhPel}. This 
simplification is the result of the different behaviour of many- and singe-particle 
distribution functions during the evolution in time. In fact, at the initial stage 
of evolution, when time $t$ is of the order of the typical time $\tau _{0} $ of the 
correlation decay (which in turn, is of duration of a single collision), many-particle 
distribution functions $f_{S} \left({x}_{1} ,...,{x}_{S} ;t\right)$ change rapidly, 
in contrast to the single-particle distribution function $f_{1} \left({x},t\right)$ 
that experiences significant changes over time periods longer than the typical relaxation 
time $\tau _{r} $ necessary to reach an equilibrium state, $\tau _{r} \gg \tau _{0} $. 
Such a difference in the evolutionary behaviour of  a single- and many-particle distribution 
functions formed the basis of the Bogolyubov's idea about hierarchy of relaxation times 
\cite{Bogolyubov}.

According to the idea of a relaxation times hierarchy, the evolution of many-particle system can be divided into several stages. The simplest scenario is as follows. At $\tau _{0} \ll t\ll \tau _{r} $ there is a kinetic stage of evolution of the system when the system behaviour can be described by a single-particle distribution function, only. Further simplification in the description of many-particle system occurs at $t\gg \tau _{r} $ (hydrodynamic stage of the evolution of the system), when the behaviour of the system can be described by hydrodynamic parameters, e.g., the particle density, average speed and temperature of the medium. In this paper, the reduced description method will be used for the derivation of the kinetic equations describing evolution of dissipative systems in the external stochastic field. Mathematically, the reduced description method introduces time-dependence
of many-particle distribution functions $f_{S} \left({x}_{1} ,...,{x}_{S} ;t\right)$ as a
functional time-dependence via the reduced description parameters at the corresponding
stage of evolution. In particular, at the kinetic stage the many-particle distribution functions depend on time only through the single-particle distribution function $f_{1} (\mathbf{x}',t)$,
\begin{equation}\label{eq__3_4_}
f_{S} \left({x}_{1} ,...,{x}_{S} ;t\right)=f_{S} \left({x}_{1} ,...,{x}_{S} ;f_{1} ({x}',t)\right).
\end{equation}
Note, that the $x'$ - dependence on the right hand side of (\ref{eq__3_4_}) implies functional dependence on the PDF $f_1$.
In addition to the functional hypothesis (\ref{eq__3_4_}), the principle of spatial correlation weakening is put into basis of the reduced description method. It can be summarized as follows. Let $S$ particles be divided into two subgroups containing $S'$ and $S''$ particles respectively $S=S'+S''$. If the distance $R$ between these subgroups of particles increases, $R\to \infty $, then due to the weakening of correlations between the particles the $S$- particle distribution function is decomposed into the product of the distribution functions related to each subgroup of particles:
\begin{equation}\label{eq__3_5_}
f_{S} \left({x}_{1} ,...,{x}_{S} ;t\right)\mathop{\longrightarrow }\limits_{R\to \infty } f_{S'} \left({x}'_{1} ,...,{x}'_{S} ;t\right)f_{S''} \left({x}''_{1} ,...,{x}''_{S} ;t\right),
\end{equation}
where the sign ``prime'' is used to denote the coordinates and momenta of the particles from the subgroup $S'$, and ``two-primes'' to denote the coordinates and momenta of the particles from
the subgroup $S''$. It should be noted, however, that the principle of spatial correlations weakening (\ref{eq__3_5_}) refers to many-particles distribution functions where the thermodynamic limit transition $V\to \infty $, $\left(N/V\right)=\mathrm{const}$ was performed.

According to (\ref{eq__3_4_}), the time derivative $\frac{\partial f_{S} }{\partial t} $ in  (\ref{eq__3_3_}) with $S\ne 1$ should be understood as follows:
\begin{equation}\label{eq__3_6_}
\frac{\partial }{\partial t} f_{S} \left({x}_{1} ,...,{x}_{S} ;f_{1} ({x},t)\right)=\int  {\rm d} {x}' \, \frac{\delta f_{S} \left({x}_{1} ,...,{x}_{S} ;f_{1} ({x},t)\right)}{\delta f_{1} ({x}',t)}  \frac{\partial f_{1} ({x}',t)}{\partial t} ,
\end{equation}
where $\frac{\delta f_{S} \left(f_{1} ({x},t)\right)}{\delta f_{1} ({x}',t)} $ is the functional derivative. According to (\ref{eq__3_3_}), the single-particle distribution function $f_{1} ({x}',t)$ satisfies the following equation:
\begin{eqnarray} \label{eq__3_7_}
\nonumber \frac{\partial f_{1} }{\partial t} +\frac{\mathbf{p}_{1} }{m} \frac{\partial f_{1} }{\partial \mathbf{x}_{1} } +\sum _{n=2}^{\infty }\frac{\left(-1\right)^{n-1} }{2^{n-1} \left(n-1\right)!}  y_{i_{1} ...i_{n} } \left(\mathbf{x}_{1} ,...,\mathbf{x}_{1} \right)\frac{\partial ^{n} f_{1} }{\partial p_{1i_{1} } ...\partial p_{1i_{n} } } \\
 =\frac{1}{v} L\left({x}_{1} ;f_{1} \right),   \qquad v\equiv \frac{V}{N} ,
\end{eqnarray}
where $L\left({x}_{1} ;f_{1} \right)$ is the (generalized) collision integral defined by the formula
\begin{equation}\label{eq__3_8_}
L\left({x}_{1} ;f_{1} \right)\equiv -\frac{\partial }{\partial \mathbf{p}_{1} } \int  {\rm d} {x}_{2} \, f_{2} \left({x}_{1} ,{x}_{2} ;f_{1} \right)\mathbf{F}_{1,2} .
\end{equation}
In order to get a closed equation for a single-particle probability density one should find 
the collision integral (\ref{eq__3_8_}) as a functional of a single-particle distribution 
function, thus, one should truncate the infinite set of equations (\ref{eq__3_3_}).
Such a truncation can be performed by using either low density or weak interaction approximation. 
In the next Section we demonstrate the cut off procedure and develop a perturbation theory in case 
of a weak interaction between the particles. We thus do not use the low density expansion and 
respectively, do not discuss the issues of divergency and non-analytical behavior of transport 
coefficients \cite{Ernst}.

\section{\label{sec.5}Kinetic equations for dissipative system of weakly interacting particles}

We demonstrate such cut off procedure and develop a perturbation theory in case of weak interaction 
between particles embedded in a weak external stochastic field. Using (\ref{eq__3_6_}), the set 
of equations (\ref{eq__3_3_}) can be written as
\begin{equation}\label{eq__3_9_}
-\int  {\rm d} {x} \, \frac{\delta f_{S} \left(f_{1} \right)}{\delta f_{1} ({x},t)} \frac{\mathbf{p}}{m}  \frac{\partial f_{1} (\mathbf{x},t)}{\partial \mathbf{x}} +\sum _{1\le \alpha \le S}\frac{\mathbf{p}_{\alpha } }{m} \frac{\partial f_{S} \left(f_{1} \right)}{\partial \mathbf{x}_{\alpha } } =\frac{1}{v} K_{S} \left(f_{1} \right) ,
\end{equation}
where
\begin{eqnarray}\label{eq__3_10_}
\nonumber K_{S} \left(f_{1} \right) && \equiv  -v\sum _{1\le \alpha <\beta \le S}\frac{\partial f_{S} \mathbf{F}_{\alpha ,\beta } }{\partial \mathbf{p}_{\alpha } }  -\sum _{1\le \alpha \le S}\frac{\partial }{\partial \mathbf{p}_{\alpha } }  \int  {\rm d} \mathbf{x}_{S+1} \, f_{S+1} \mathbf{F}_{\alpha ,S+1} \\
\nonumber && -\int  {\rm d} \mathbf{x} \, \frac{\delta f_{S} \left(f_{1} \right)}{\delta f_{1} ({x},t)}  \left\{L\left({x};f_{1} \right)-v\sum _{n=2}^{\infty }\frac{\left(-1\right)^{n-1} }{2^{n-1} \left(n-1\right)!}  y_{i_{1} ...i_{n} } \left(\mathbf{x},...,\mathbf{x}\right)\right.\\
\nonumber && \times \left.  \frac{\partial ^{n} f_{1} }{\partial p_{i_{1} } ...\partial p_{i_{n} } } \right\}- \sum _{n=2}^{\infty }\frac{\left(-1\right)^{n-1} }{2^{n-1} \left(n-1\right)!}\\
&& \times  \sum _{1\le \alpha _{1} \le S}... \sum _{1\le \alpha _{n} \le S}y_{i_{1} ...i_{n} } \left(\mathbf{x}_{\alpha _{1} } ,...,\mathbf{x}_{\alpha _{n} } \right) \frac{\partial ^{n} f_{S} }{\partial p_{\alpha _{1} i_{1} } ...\partial p_{\alpha _{n} i_{n} } } .
\end{eqnarray}
The chain (\ref{eq__3_9_}) should be supplied with ``initial conditions''. For those, we use the fact that many-particle distribution functions satisfy the principle of spatial correlation weakening, (\ref{eq__3_5_}). To this end, following \cite{Bogolyubov,AkhPel}, we introduce the auxiliary parameter $\tau $, having the dimension of time, but not necessarily representing the physical time. We next consider the many-particle distribution function $f_{S} \left(\mathbf{x}_{1} -\frac{\mathbf{p}_{1} }{m} \tau ,\mathbf{p}_{1} ,...,\mathbf{x}_{S} -\frac{\mathbf{p}_{S} }{m} \tau ,\mathbf{p}_{S} ;f_{1} \right)$. According to (\ref{eq__3_5_}) this function must satisfy the asymptotic relation:
\begin{equation}\label{eq__3_11_}
f_{S} \left(\mathbf{x}_{1} -\frac{\mathbf{p}_{1} }{m} \tau ,\mathbf{p}_{1} ,...,\mathbf{x}_{S} -\frac{\mathbf{p}_{S} }{m} \tau ,\mathbf{p}_{S} ;f_{1} \right)\mathop{\longrightarrow }\limits_{\tau \to \pm\infty } \prod _{1\le \alpha \le S}f_{1} \left(\mathbf{x}_{\alpha } -\frac{\mathbf{p}_{\alpha } }{m} \tau ,\mathbf{p}_{\alpha } \right).
\end{equation}
If we further define the shift operator $\hat{\Lambda }_{S}^{0} $ in coordinate space by a formula
\begin{equation}\label{eq__3_12_}
i\hat{\Lambda }_{S}^{0} \equiv \sum _{1\le \alpha \le S}\frac{\mathbf{p}_{\alpha } }{m} \frac{\partial }{\partial \mathbf{x}_{\alpha } }  ,
\end{equation}
than (\ref{eq__3_11_}) can be rewritten as:
\begin{equation}\label{eq__3_13_}
e^{i\tau \hat{\Lambda }_{S}^{0} } f_{S} \left(\tau \right)\mathop{\longrightarrow }\limits_{\tau \to \pm\infty } \prod _{1\le \alpha \le S}f_{1} \left({x}_{\alpha } \right) ,
\end{equation}
where $\exp \left(i\tau \hat{\Lambda }_{S}^{0} \right)$ is a so-called ``free evolution operator'' and
\begin{eqnarray} \label{eq__3_14_}
\nonumber f_{S} \left(\tau \right) && \equiv f_{S} \left({x}_{1} ,...,{x}_{S} ;e^{-i\tau \hat{\Lambda }_{1}^{0} } f_{1} \left({x}'\right)\right)\\
&& =f_{S} \left({x}_{1} ,...,{x}_{S} ;f_{1} \left(\mathbf{x}'-\frac{\mathbf{p}'}{m} \tau ,\mathbf{p}'\right)\right).
\end{eqnarray}
With the use of the operators introduced, expression (\ref{eq__3_9_}) can be rewritten as:
\begin{equation}\label{eq__3_15_}
\frac{\partial }{\partial \tau } e^{i\tau \hat{\Lambda }_{S}^{0} } f_{S} \left(\tau \right)=\frac{1}{v} e^{i\tau \hat{\Lambda }_{S}^{0} } K_{S} \left(\tau \right),
\end{equation}
where
\begin{eqnarray} \label{eq__3_16_}
\nonumber K_{S} \left(\tau \right) && \equiv K_{S} \left({x}_{1} ,...,{x}_{S} ;e^{-i\tau \hat{\Lambda }_{1}^{0} } f_{1} \left({x}'\right)\right)\\
&& =K_{S} \left({x}_{1} ,...,{x}_{S} ;f_{1} \left(\mathbf{x}'-\frac{\mathbf{p}'}{m} \tau ,\mathbf{p}'\right)\right).
\end{eqnarray}
Integrating (\ref{eq__3_15_}) over $\tau $ between $-\infty $ and 0 and using the asymptotic conditions (\ref{eq__3_13_}), we obtain
\begin{equation}\label{eq__3_17_}
f_{S} \left({x}_{1} ,...,{x}_{S} ;f_{1} \left({x}'\right)\right)=\prod _{1\le \alpha \le S}f_{1} \left({x}_{\alpha } \right) +\frac{1}{v} \int _{-\infty }^{0}{\rm d}\tau \, e^{i\tau \hat{\Lambda }_{S}^{0} } K_{S} \left(\tau \right).
\end{equation}
Equation (\ref{eq__3_17_}) demonstrates that the assumption of factorization
of many-particle distribution function is valid only if the
second term in the right-hand side is negligibly small. On the
other hand, Eq.~(\ref{eq__3_17_}) allows us to develop a perturbation theory.
Indeed, it follows from Eqs.~(\ref{eq__3_10_}) and (\ref{eq__3_8_}) that the second term
in the right-hand side of Eq.~(\ref{eq__3_17_}) is small if the intensity of the
stochastic field and the particle interactions (potential and
dissipative ones) are weak enough. Under these assumptions, the value $K_{S} \left(\tau \right)$, see (\ref{eq__3_10_}), can be considered 
as small and, therefore, in the main approximation we have
\begin{equation*}
f_{S} \left({x}_{1} ,...,{x}_{S} ;f_{1} \left({x}'\right)\right)=\prod _{1\le \alpha \le S}f_{1} \left({x}_{\alpha } \right) ,
\end{equation*}
which implies
\begin{equation}\label{eq__3_18_}
f_{2} \left({x}_{1} ,{x}_{2} \right)=f_{1} \left({x}_{1} \right)f_{1} \left({x}_{2} \right).
\end{equation}
Next, substituting (\ref{eq__3_18_}) into (\ref{eq__3_8_}) and using (\ref{eq__1_5_}) and (\ref{eq__3_7_}), we obtain the following closed kinetic equation,
\begin{eqnarray}\label{eq__3_19_}
\nonumber \frac{\partial f_{1} }{\partial t}  +\frac{\mathbf{p}_{1} }{m} \frac{\partial f_{1} }{\partial \mathbf{x}_{1} } && +\sum _{n=2}^{\infty }\frac{\left(-1\right)^{n-1} }{2^{n-1} \left(n-1\right)!}  y_{i_{1} ...i_{n} } \left(\mathbf{x}_{1} ,...,\mathbf{x}_{1} \right)\frac{\partial ^{n} f_{1} }{\partial p_{1i_{1} } ...\partial p_{1i_{n} } }\\
&& =\frac{1}{v} \frac{\partial }{\partial \mathbf{p}_{1} } f_{1} \left({x}_{1} \right)\left(\int  {\rm d} {x}_{2} \, f_{1} \left({x}_{2} \right)\frac{\partial V_{1,2} }{\partial \mathbf{x}_{1} } +\int  {\rm d} {x}_{2} \, f_{1} \left({x}_{2} \right)\frac{\partial R_{1,2} }{\partial \mathbf{p}_{1} } \right),
\end{eqnarray}
which describes the kinetic stage evolution of many-particle dissipative systems under 
the influence of an external stochastic field. If we make a simplest assumption 
that the dissipation function is quadratic in the particles' momenta, 
then the function $R_{1,2} $, following \cite{Landau}, can be chosen as
\begin{eqnarray}\label{eq__3_20_}
\nonumber R_{1,2} =\frac{1}{2} \tilde{\gamma }\left(\mathbf{x}_{1} -\mathbf{x}_{2} \right)\left(\mathbf{p}_{1} -\mathbf{p}_{2} \right)^{2}, \\
\tilde{\gamma }\left(\mathbf{x}_{1} -\mathbf{x}_{2} \right)>0.
\end{eqnarray}
If there is no dissipation, $\tilde{\gamma }\left(\mathbf{x}_{1} -\mathbf{x}_{2} \right)=0$, (\ref{eq__3_19_}) is reduced to the Vlasov equation for a system in external stochastic field, with the self-consistent field
\begin{equation}\label{eq__3_21_}
U\left(\mathbf{x}_{1} \right)=\frac{1}{v} \int  {\rm d} \mathbf{x}_{2} \, V\left(\mathbf{x}_{1} -\mathbf{x}_{2} 
\right)\int {\rm d}\mathbf{p}_{2} \, f_{1} \left(\mathbf{x}_{2} , \mathbf{p}_{2} \right). 
\end{equation}
In general case, equation (\ref{eq__3_19_}) can be treated as the Fokker-Planck equation for the system 
of weakly interacting particles influenced by non-Gaussian stochastic field. We dwell 
on this more in the next Section.

\section{\label{sec.6}Two particular cases of the Fokker-Planck equation}

\subsection{Gaussian stochastic field}

We consider first (\ref{eq__3_19_}) with the dissipation function given
by (\ref{eq__3_20_}), and assume the Gaussian statistics of the external stochastic field. Thus, the odd 
correlation functions of the field vanish, and the even ones are expressed via the pair correlation function. 
We choose it in the form $y_{ij} \left(\mathbf{x}_{1} ,\mathbf{x}_{2} \right)\equiv 
\delta _{ij} g\left(\mathbf{x}_{1} ,\mathbf{x}_{2} \right)$  to simplify calculations. 
Also for simplicity, we assume that the single-particle distribution function is isotropic in the momentum space,
\begin{equation}\label{eq__3_22_}
f_{1} \left(\mathbf{x},\mathbf{p},t\right)\equiv f_{1} \left(\mathbf{x},\left|\mathbf{p}\right|,t\right). 
\end{equation}
Equation (\ref{eq__3_19_}) takes then the ``traditional'' form of the Fokker - Planck equation,
\begin{eqnarray} \label{eq__3_23_}
\nonumber \frac{\partial f_{1} \left({x},t\right)}{\partial t} && +\frac{\mathbf{p}}{m} 
\frac{\partial f_{1} \left({x},t\right)}{\partial \mathbf{x}} -\frac{\partial U\left(\mathbf{x}\right)}
{\partial \mathbf{x}} \frac{\partial f_{1} \left(\mathbf{x},t\right)}{\partial \mathbf{p}}\\
&& =\frac{\partial }{\partial \mathbf{p}} \left(\gamma \left(\mathbf{x},t\right)\mathbf{p}f_{1} 
\left({x},t\right)+\frac{1}{2} g\left(\mathbf{x},\mathbf{x}\right)\frac{\partial f_{1} \left({x},t\right)}
{\partial \mathbf{p}} \right),
\end{eqnarray}
where the self-consistent field $U\left(\mathbf{x},t\right)$ is given by (\ref{eq__3_21_}), 
and  $\gamma \left(\mathbf{x},t\right)$ is determined by
\begin{equation}\label{eq__3_24_}
\gamma \left(\mathbf{x},t\right)\equiv \frac{1}{v} \int  {\rm d} \mathbf{x}' \,
\tilde{\gamma }\left(\mathbf{x}-\mathbf{x}'\right)\int {\rm d}\mathbf{p}' \, f_{1} \left({x}',t\right), 
\qquad \gamma \left(\mathbf{x}\right)>0 . 
\end{equation}

\subsection{Homogeneous system embedded in non-Gaussian stochastic field}
Let us now consider spatially homogeneous system. In this case, the correlation functions 
$y_{i_{1} ...i_{n} } \left(\mathbf{x}_{\alpha _{1} } ,...,\mathbf{x}_{\alpha _{n} } \right)$, 
see (\ref{correlation})
and (\ref{eq__3_10_}), depend only on the coordinate difference $\mathbf{x}_{\alpha _{i} } -\mathbf{x}_{\alpha _{j} } $. 
Then, the correlation functions $y_{i_{1} ...i_{n} } \left(\mathbf{x}_{1} ,...,\mathbf{x}_{1} \right)$ 
entering (\ref{eq__3_19_}) do not depend on the spatial coordinate, 
\begin{equation}\label{eq__3_25_}
y_{i_{1} ...i_{n} } \left(\mathbf{x}_{1} ,...,\mathbf{x}_{1} \right)\equiv y_{i_{1} ...i_{n} } .
\end{equation}
Thus, in the spatially homogeneous case (\ref{eq__3_19_}) takes the form (assuming that (\ref{eq__3_20_}) is also valid) 
\begin{eqnarray}\label{eq__3_26_}
\nonumber \frac{\partial f_{1} \left(\mathbf{p},t\right)}{\partial t} && +\sum _{n=2}^{\infty }
\frac{\left(-1\right)^{n-1} }{2^{n-1} \left(n-1\right)!}  y_{i_{1} ...i_{n} } 
\frac{\partial ^{n} f_{1} \left(\mathbf{p},t\right)}{\partial {p}_{i_{1} } ...\partial {p}_{i_{n} } } \\
&& =\frac{\tilde{\gamma }}{v} \frac{\partial }{\partial \mathbf{p}_{1} } f_{1} 
\left(\mathbf{p}_{1} \right)\int {\rm d}\mathbf{p}_{2} \, f_{1} \left(\mathbf{p}_{2} 
\right)\left(\mathbf{p}_{1} -\mathbf{p}_{2} \right),
\end{eqnarray}
\begin{equation*}
\tilde{\gamma }\equiv \frac{1}{v} \int  {\rm d} \mathbf{x}' \, \tilde{\gamma }\left(\mathbf{x}-\mathbf{x}'\right).
\end{equation*}
Note that the system in the momentum space may not be isotropic, in contrast to (\ref{eq__3_22_}). 
We now consider the mean kinetic energy $\bar{\varepsilon }\left(t\right)$ of the system,
\begin{equation}\label{eq__3_27_}
\bar{\varepsilon }\left(t\right)\equiv \int {\rm d}\mathbf{p} \, f_{1}  \left(\mathbf{p},t\right)\frac{\mathbf{p}^{2} }{2m} .
\end{equation}
The evolution equation for this quantity is an immediate consequence of the kinetic equation (\ref{eq__3_26_}):
\begin{equation}\label{eq__3_28_}
\frac{\partial \bar{\varepsilon }\left(t\right)}{\partial t} =\frac{3}{2} \frac{y_{jj} \rho}{m} -\frac{\tilde{\gamma }}{2mv} \int {\rm d}\mathbf{p}_{1}  \int {\rm d}\mathbf{p}_{2} \, f_{1} \left(\mathbf{p}_{2} ,t\right)f_{1} \left(\mathbf{p}_{1} ,t\right)\left(\mathbf{p}_{1} -\mathbf{p}_{2} \right)^{2} ,
\end{equation}
where $\rho$ is the particle density,
\begin{equation}\label{eq__3_29_}
\rho\equiv \int {\rm d}\mathbf{p} \, f_{1} \left(\mathbf{p},t\right),
\end{equation}
and the following equalities hold,
\begin{equation}\label{eq__3_30_}
\frac{\partial }{\partial t} \int {\rm d}\mathbf{p} \, f_{1} \left(\mathbf{p},t\right)=0, \qquad \frac{\partial }
{\partial t} \int {\rm d}\mathbf{p} \, f_{1} \left(\mathbf{p},t\right)\mathbf{p}=0,
\end{equation}
which imply conservation of the density and momentum.

From (\ref{eq__3_28_}) one may conclude that the investigated system can be heated or cooled, 
depending on the sign of the right hand side of (\ref{eq__3_28_}). If we assume that initially 
the average momentum of the particles equals zero, then according to (\ref{eq__3_30_}), it is 
always equal to zero, and (\ref{eq__3_28_}) takes the form
\begin{equation}\label{eq__3_31_}
\frac{\partial \bar{\varepsilon }\left(t\right)}{\partial t} =-2\gamma \bar{\varepsilon }
\left(t\right)+\frac{3}{2} \frac{y_{jj} \rho}{m} , \gamma \equiv \rho\tilde{\gamma },
\end{equation}
where we take into account that $\rho$ is constant, and therefore, the coefficient $\gamma $ 
does not depend on time. Taking $\bar{\varepsilon }\left(t=0\right) = \bar{\varepsilon }_{0}$, 
we get the solution of (\ref{eq__3_31_}),
\begin{equation}\label{eq__3_32_}
\bar{\varepsilon }\left(t\right)=\frac{3}{4} \frac{y_{jj} \rho}{m\gamma } +\left(\bar{\varepsilon }_{0} 
-\frac{3}{4} \frac{y_{jj} \rho}{m\gamma } \right)\exp \left(-2\gamma t\right),
\end{equation}
From this equation one can see that the mean kinetic energy of the system decreases during 
the evolution (the system cools down) if $\bar{\varepsilon }_{0} >\frac{3}{4} \frac{y_{jj} \rho}{m\gamma } $, 
while the system is heated, if $\bar{\varepsilon }_{0} <\frac{3}{4} \frac{y_{jj} \rho}{m\gamma } $. 
If $\bar{\varepsilon }_{0} =\frac{3}{4} \frac{y_{jj} \rho}{m\gamma } $, the mean kinetic energy 
does not change (we note that the homogeneous cooling state of a granular flow in the low-density 
limit was investigated in~\cite{brey1996, brey1997, brey2004}). As it is seen from (\ref{eq__3_32_}), the applied
stochastic force can reverse the sign of the effect, leading to the heating of dissipative system. 
If the Maxwell distribution is established during the evolution of the gaseous dissipative system, 
the expression for the equilibrium mean kinetic energy $\bar{\varepsilon }\left(\infty \right)=
\frac{3}{4} \frac{y_{jj} \rho}{m\gamma } $, that follows from (\ref{eq__3_32_}), coincides with 
the well-known formula $\bar{\varepsilon }\left(\infty \right)=\frac{3}{2} \rho T$, where  $T$ 
is the temperature of the medium. That means that the pair correlation function of the external 
stochastic field $y_{jj} $ satisfies the relation $y_{jj} =2m\gamma T$, known in the theory 
of classical Brownian motion. 

We note that for the non-interacting Brownian particles the concepts of equilibrium, 
Maxwell distribution and the temperature of a heat bath have a well-known physical meaning. 
For the dissipative systems, as far as the authors know, to define a notion of temperature
is a non-trivial issue. A relevant discussion can be 
found in \cite{Puglisi2005, Gold2008a, Gold2008b}. 

We also note that according to (\ref{eq__3_28_}) and (\ref{eq__3_32_}), only 
pair correlation function of external field is responsible for pumping energy 
into the system. Therefore, (\ref{eq__3_28_}) and (\ref{eq__3_31_}) also hold 
in case of the Gaussian external field, see (\ref{eq__3_23_}). Obviously, 
this is because the
kinetic energy of a single particle depends on the square of the momentum, 
$\varepsilon _{p} ={p^{2} \mathord{\left/ {\vphantom {p^{2}  2m}} \right. \kern-\nulldelimiterspace} 2m} $, 
see (\ref{eq__3_26_}) and (\ref{eq__3_27_}). The higher order correlation functions 
are involved in the evolution equations for higher order moments of the single-particle 
distribution function.

In this Section we considered only the simplest example of the mean energy evolution. 
Indeed, (\ref{eq__3_26_}), (\ref{eq__3_28_}), (\ref{eq__3_31_}) and (\ref{eq__3_32_}) were 
obtained for a special form (\ref{eq__3_20_}) of the dissipative function $R_{1,2} $, and 
with additional assumption of homogeneity, see (\ref{eq__3_25_}). In a more general case, 
as follows from (\ref{eq__3_19_}) and the definition (\ref{eq__3_27_}), the mean energy 
evolution can be more complex than predicted by (\ref{eq__3_32_}) or (\ref{eq__3_28_}).

\section{\label{sec.7}Summary}
In this paper, we contributed to developing the kinetic theory of many-particle dissipative systems 
in an external stochastic field. We suggested a procedure to obtain 
an infinite BBGKY hierarchy 
for many-particle distribution functions by using the Furutsu-Novikov formula generalized 
to the case of a non-Gaussian stochastic field, provided it has moments of any order. Further, 
we extended the reduced description method and developed a technique to derive
a kinetic 
equation in the case of a weak interaction between the particles and a 
low intensity of the external 
stochastic field. In the case of a Gaussian external field the kinetic equation becomes 
a Fokker-Planck equation in its ``standard'' form, with the friction 
coefficient expressed through
the one-particle distribution function. As another simple particular case, we considered 
the mean energy evolution in a homogeneous system demonstrating competition between 
the processes of cooling and heating. 

We now briefly outline the direct extensions of the perturbation theory developed. 
First, it allows one to reduce the assumptions of weak interaction 
between the particles and weak intensity of the external field by calculating higher order 
corrections to the kinetic equation. Second, a similar approach can be used to construct 
the kinetic theory in the case of a low particle density, 
with no assumption of a weak particle interaction (as long as this interaction does not lead 
to the formation of bound states). Along these lines, the Boltzmann equation for dissipative randomly
driven system arises (along with the problem of the divergence of
higher order density corrections; however
it does not rule out the significance of the Boltzmann equation). 
Also, an interesting possibility within the framework of the reduced description method is to take 
into account long-lived fluctuations which are of importance for dissipative systems. 
In fact, the reduced description method is applicable to systems with 
long lasting memory effects, however, it requires modifications as mentioned
in Ref.\cite{Yura},
where the equations of fluctuating hydrodynamics have been derived, together 
with reproducing the 
results for long time tails. 
We believe that this theory can straightforwardly be extended to
dissipative systems. Similar procedures generalizing Bogolyubov's theory have 
been used in Refs.\cite{Krommes} and \cite{Klimontovich}.

\section*{Acknowledgments}
AVCh thanks Berlin Mathematical Society for financial support.


\begin{thebibliography}{00}

\bibitem{Bogolyubov} N. Bogolyubov \textit{Problems of Dynamical Theory in Statistical Physics} (Providence, RI, Providence College, 1959)

\bibitem{AkhPel} A.I. Akhiezer, S.V. Peletminskii \textit{Methods of Statistical Physics} (Oxford, Pergamon, 1981)

\bibitem{Uhlenbeck} G.E. Uhlenbeck, G.W. Ford \textit{Lectures in Statistical Mechanics} (Providence, American Mathematical Society, 1963)

\bibitem{Ernst} M.H. Ernst In: W. Sung, I. Chang, K. Byungnam et al (Eds) \textit{Progress in Statistical Physics} (Singapore,
World Scientific, 1998), PP. 3-28

\bibitem{Cohen50} E.G.D. Cohen, {Physica A} \textbf{194}, 229 (1993).

\bibitem{klim1} Yu.L. Klimontovich \textit{The Kinetic Theory of Electromagnetic Processes} (Berlin, Springer-Verlag, 1983)

\bibitem{Russel} W.B. Russel, D.A. Saville and W.R. Schowalter \textit{Colloidal Dispersions}, (Cambridge, Cambridge University Press, 1989)

\bibitem{Edwards} M. Doi, S.F. Edwards \textit{The Theory of Polymer Dynamics} (Oxford, Clarendon Press, 1986)

\bibitem{Pokrovsky} V.N. Pokrovsky \textit{The Mesoscopic Theory of Polymer Dynamics} (Springer Science + Business Media: 
Springer Series in Chemical Physics 95, 2010)

\bibitem{Rubi}  M. Mayorga, L. Romero-Salazar and J.M. Rubi, {Physica A} \textbf{307}, 297 (2002).

\bibitem{Archer} A.J. Archer, M. Rauscher, {J. Phys. A: Math. Gen.} \textbf{37}, 9325 (2004).

\bibitem{Basu} A. Basu, S. Ramaswamy, {J. Stat. Mech.} \textbf{11}, 11003 (2007).

\bibitem{HuiXia} H. Xia , K. Ishii, T. Iwaii, H. Li and B. Yang, {Applied Optics} \textbf{47}, 1257 (2008).

\bibitem{Chavanis} P.-H. Chavanis, {Physica A} \textbf{390}, 1546 (2011).

\bibitem{Brilliantov} N.V. Brilliantov, T. Poschel \textit{Kinetic Theory of Granular Gases} (Oxford, Oxford University Press, 2004) 

\bibitem{Mehta}  A. Mehta \textit{Granular Physics} (Cambridge, Cambridge University Press, 2007)

\bibitem{PRreview} D. Chowdhury, L. Santen, A. Schadschneider, {Phys. Rep.} \textbf{329}, 199 (2000).

\bibitem{RMPreview} D. Helbing, {Rev. Mod. Phys} \textbf{73}, 1067 (2001).

\bibitem{Barrat} A. Barrat, E. Trizac, M.H. Ernst, {J. Phys.: Condens. Matter} \textbf{17}, S2429 (2005).

\bibitem{Noije1998} T.P.C. Van Noije, M.H. Ernst, R. Brito, {Physica A} \textbf{251}, 266 (1998).

\bibitem{Noije1999} T.P.C. Van Noije, M.H. Ernst, E. Trizac, I. Pagonabarraga, {Phys. Rev. E} \textbf{59}, 4326 (1999).

\bibitem{Ernst2000} M.H. Ernst In: J. Karkheck (Ed) {Dynamics: Models and Kinetic Methods for Non-Equilibrium
Many Body Systems.} Book Series: NATO Advanced Science Institutes Series, Series E, Applied Sciences \textbf{371}, 239 (2000).

\bibitem{Noije2001} T.P.C. Noije, M.H. Ernst In: T. Poschel, S. Luding (Eds) {Granular gases.} Book Series:
Lecture Notes in Physics \textbf{564}, 3 (2001).

\bibitem{Pag} I. Pagonabarraga, E. Trizac, T.P.C. van Noije, M.H. Ernst, {Phys. Rev.} E \textbf{65}, 011303 (2001).

\bibitem{Maynar} P. Maynar, M.I.G. de Soria, E. Trizac, {Eur. Phys. J. Special Topics} \textbf{179}, 123 (2009).

\bibitem{Maynar_2} P. Maynar, M.I. Garcia de Soria, {Math. Model. Nat. Phenom.} \textbf{6}, 87 (2011).

\bibitem{Khalil} N. Khalil, Garz\'o, {Phys. Rev.} E \textbf{88}, 052201 (2013).

\bibitem{Soria} M.I.G. De Soria, P. Maynar, E. Trizac, {Phys. Rev. E} \textbf{87}, 022201 (2013). 

\bibitem{Prados} A. Prados, E. Trizac, {Phys. Rev. Lett.} \textbf{112}, 198001 (2014).

\bibitem {Rama} S. Ramaswamy, {Annu. Rev. Condens. Matter Phys.} \textbf{1}, 323 (2010).

\bibitem{Pavel} P. Romanczuk, M. Bar, W. Ebeling, B. Lindner and L. Schimansky-Geier, {Eur. Phys. J. Special Topics} \textbf{202}, 1--162 (2012).

\bibitem{Marchetti} M.C. Marchetti, J.F. Joanny, S. Ramaswamy, T.B. Liverpool, J. Prost, M.
Rao and R. Aditi Simha, {Rev. Mod. Phys.} \textbf{85}, 1143 (2013).

\bibitem{Pavel2} R. Grossman, L. Schimansky-Geier, P. Romanczuk, {New J. Phys.} \textbf{15}, 085014 (2013).

\bibitem{Ihle} T. Ihle, {Eur. Phys. J. Special Topics} \textbf{223}, 1293 (2014).

\bibitem{Peshkov} A. Peshkov, E. Bertin, F. Ginelli, H. Chat\'e, {Eur. Phys. J. Special Topics} \textbf{223}, 1315 (2014).

\bibitem{Peshkov2} E. Bertin, H. Chat\'e, F. Ginelli, G. Gr\'egoire, S. L\'eonardo, A. Peshkov, {Eur. Phys. J. Special Topics}
\textbf{223}, 1419 (2014).

\bibitem{Ihle2} T. Ihle, {Eur. Phys. J. Special Topics} \textbf{223}, 1423 (2014).

\bibitem{Laskin} N.V. Laskin, S.V. Peletminskii and V.I. Prikhod'ko, {Journal of Physical Studies} \textbf{2}, 3 (1998).

\bibitem{Archive}  I. Goldhirsch, S.V. Peletminskii, A.S. Peletminskii and A.I. Sokolovskii {arXiv}: 1307.3466v1 [cond-mat.stat-mech] (2013)

\bibitem{Furutsu} K. Furutsu, {J. Res. N.B.S.} \textbf{D-67}, 303 (1963).

\bibitem{Novikov} E.A. Novikov, {Zh. Eksp. Teor. Fiz.} \textbf{47} (1919) (Sov. Phys. {JETP} 1964).

\bibitem{nonGaussian} S.V. Peletminskii, Yu.V. Slyusarenko and A.I. Sokolovskii, {Physica} A \textbf{326}, 412 (2003).

\bibitem{Landau} L.D.  Landau, E.M. Lifshitz \textit{Statistical physics (Course of theoretical physics, volume 5)} (London, Oxford, 1980)

\bibitem{brey1996} J.J. Brey, M.J. Ruiz-Montero and D. Cubero, {Phys. Rev.} E \textbf{54}, 3664 (1996).

\bibitem{brey1997} J.J. Brey, J.W. Dufty and A. Santos, {J. Stat. Phys.} \textbf{87}, 1051 (1997).

\bibitem{brey2004} J.J. Brey, M.I. Garcia de Soria, P. Maynar and M.J. Ruiz-Montero, {Phys. Rev. }E \textbf{70}, 011302 (2004).

\bibitem{Puglisi2005} A. Baldassarri, A. Barrat, G. D'Anna, V. Loreto, P. Mayor, A. Puglisi, {J. Phys.: Condens. Matter} \textbf{17}, S2405 (2005).

\bibitem{Gold2008a} I. Goldhirsch, {Powder Technology} \textbf{182}, 130 (2008).

\bibitem{Gold2008b}  D. Serero, C. Goldenberg, S.H. Noskowicz and I. Goldhirsch, {Powder Technology} \textbf{182}, 257 (2008).

\bibitem{Yura} S. Peletminsky, Y. Slusarenko, {Physica} A \textbf{210}, 165 (1994).

\bibitem{Krommes} J.A. Krommes, C. Oberman, {J. Plasma Phys.} \textbf{16}, 193 (1976).

\bibitem{Klimontovich} Yu.L. Klimontovich, \textit{Statistical Theory of Open Systems, vol.1.} (Dordrecht, Kluwer Academic Publishers, 1995)




\end{thebibliography}
\end{document}